\documentclass{formato_generico}

\title{Nonlinear Classical Dynamics described by a Density Matrix in the Classical Limit}

\author{Gaspar Gonzalez$^{1,2}$ \, Angelo Plastino$^{3}$ \, Andrés Kowalski$^{1,2}$}

\affiliation{%
$^{1}$ Instituto de Física La Plata (IFLP-CONICET/UNLP), La Plata,  Buenos Aires, Argentina\\
$^{2}$ Comisión de Investigaciones Científicas (CIC), La Plata, 1900, Buenos Aires, Argentina\\
$^{3}$ Universidad Nacional de La Plata, La Plata,  Buenos Aires, Argentina
}
\email{kowalski@fisica.unlp.edu.ar}

\begin{document}
\maketitle

\begin{abstract}
We examine the classical limit of a fairly general nonlinear semiclassical hybrid system within a MaxEnt framework. The consistency of the hybrid dynamics requires algebraic constraints on quantum operators and smoothness conditions for the classical variables.
Analytically, we demonstrate that the classical limit is characterized by a pure density matrix representing a single state, which reproduces the dynamics of its classical analogue. To illustrate the methodology, we revisit and synthesize two previously studied examples.
\end{abstract}


\section{Introduction}
\label{sec:intro}

The classical-quantum transition is a significant topic in physics. In this context, the classical limit of quantum mechanics (CLQM) remains a frontier issue~\cite{Bloch,Milonni,meystre07,Ring,k95}. As is well known, the prevailing view of this transition relies on the concept of decoherence, whereby the entanglement of a quantum system vanishes because of its interaction with the environment. A subtopic within this framework is the classical limit of quantum dynamics (CL).

To study the CL, we employ a semiclassical hybrid methodology in which quantum and classical variables interact \cite{k95}. 

This method is valid as an approximation when the quantum effects of one subsystem are sufficiently small with respect to the other. Historically, this approach has been widely used. Examples include the Bloch equations~\cite{Bloch}, the Jaynes–Cummings semiclassical model~\cite{Milonni,meystre07}, collective nuclear motion~\cite{Ring}, and, more generally, the interaction of matter with fields.

However, it can be considered an appropriate framework to describe the interaction of a quantum system with a classical one, for instance when a quantum system interacts with a macroscopic system during a measurement process.
Currently, mesoscopic physics, closely related to the classical limit, is a subdiscipline of condensed matter physics that focuses on intermediate-scale systems. For example, when a macroscopic electronic device is reduced to mesoscopic dimensions, it begins to exhibit quantum behavior~\cite{das10meso}. Devices used in nanotechnology are notable illustrations of mesoscopic systems~\cite{yu06nanotec}. In this domain, hybrid systems can play a significant role.

Other approaches include effective potentials~\cite{patta92}, mixed quantum–semiclassical simulations (MQS)~\cite{gonzalez2023mixed}, and related methods. Most semiclassical methodologies treat classical variables as time-dependent external fields~\cite{Milonni,meystre07}. In contrast, our method allows for mutual interaction between classical and quantum variables. Unlike these approximations, our methodology preserves the quantum rules of the sub-quantal system and the symplectic structure of the classical variables. Recent works have addressed key aspects of this hybrid paradigm. Fratino et al. \cite{Fratino14} investigated the dynamics of entanglement in a model consisting of two quantum bits coupled to a classical oscillator, revealing how quantum correlations evolve under hybrid interactions. Alonso et al. \cite{Alonso20} proposed a consistent definition of entropy and a canonical ensemble for hybrid systems, integrating classical and quantum statistical principles through a maximum entropy approach. More recently, Boghiu et al. \cite{boghiu24} explored control problems in hybrid systems.

Together, these contributions underscore the theoretical richness and interdisciplinary relevance of quantum-classical hybrid models, paving the way for novel insights into quantum technologies and the classical-quantum interface.

On the other hand, the source~\cite{kurizki2015quantum} provides a comprehensive overview of hybrid quantum systems, highlighting their ability to combine distinct physical components, such as atoms, photons, spins, and superconducting circuits, to exploit complementary advantages. The article discusses both theoretical motivations and early experimental implementations.

In our  methodology, the dynamical equations for the quantum operators are canonical, while the classical variables obey Hamilton's equations, with the generator of temporal evolution given by the expectation value of the full Hamiltonian~\cite{Kowalski2002}. This idea was previously explored in Refs.~\cite{cooper94} and~\cite{bonilla92}, although different mathematical tools were employed.

We have studied the CL by varying a relative energy parameter associated with the uncertainty principle and by comparing the hybrid quantum-classical dynamics with its classical counterpart. Using dynamical tools, we have analyzed various systems—both conservative and dissipative—revealing notable features: (1) the classical limit does exist, and (2) the quantum–classical transition exhibits three distinct zones: a quasi-quantal region, a transitional (or mesoscopic) region, and a classical region~\cite{energia}.

Subsequently, we examined these regions using information-theoretic quantifiers such as Shannon, Tsallis, and Rényi entropies, their corresponding statistical complexities, and Fisher information~\cite{BP,k19renyi,k11fisher}. All analyses were performed numerically.

We also studied the CL using a suitable MaxEnt density operator. The behavior along the path to the classical limit was analyzed numerically in Ref.~\cite{previo}, and analytically in Ref.~\cite{caosclasico}.

In the present work, we extend the analytical results of the latter reference. We study the classical limit of a general system comprising a fully general classical subsystem interacting with a quantum one. The quantum operators in the Hamiltonian are required to satisfy a Lie algebra. We find that for this system, the classical limit is described by a pure density matrix.

Our aim is to demonstrate that these features are of a general nature, representing a meaningful contribution to the understanding of quantum dynamics. The methodology can be applied to various Hamiltonians across different fields, including quantum computing, quantum information, condensed matter, and quantum optics.

Additionally, we present a synthesis of two previously studied systems in order to clarify the procedure.

The paper is organized as follows. In Sect. \ref{model}, we describe our general model and review its main features, such as the conservation of quantum properties in the quantum subsystem.

In Sect. \ref{Rhogeneral}, we introduce the general MaxEnt density operator for a hybrid Hamiltonian.

In Sect. \ref{semihybryhamil}, we present the specific Hamiltonian studied in this work. Except for certain restrictions on the quantum operators, it is general.

In Sect. \ref{maxentdenshh}, we analyze the MaxEnt density operator corresponding to our hybrid Hamiltonian.

A summary of previous studies on the classical limit is provided in Sect. \ref{previousres}. We present results for two Hamiltonians, focusing on the numerically obtained regions along the path to the classical limit.

In Sect. \ref{currentres}, we present the main results of this article regarding the classical limit. We analytically derive a pure density matrix representing a well-defined state that evolves classically in time.

Conclusions are drawn in Sect. \ref{Concl}.

Finally, two appendices are included.  Appendix \ref{AppeC} describes the continuous dependence of dynamical solutions on the initial conditions, and Appendix \ref{AppeA} presents a functional form based on the unitary transformation operator used in this work.

\section{Quantum-Classical Hybrid Model}
\label{model}

We consider a Hamiltonian operator $\hat{H}$ that couples classical degrees of freedom to a quantum subsystem:
\begin{equation}
\hat{H}
= \hat{H}_{q} \;+\; H_{\mathrm{cl}}\,\hat{I}
\;+\;\hat{H}_{\mathrm{cl}}^{q}\,,
\label{Hgen}
\end{equation}
where $\hat{H}_{q}$ is the purely quantum Hamiltonian, $H_{\mathrm{cl}}$ is the classical Hamiltonian, and $\hat{H}_{\mathrm{cl}}^{q}$ describes their interaction. The operator $\hat{I}$ is the identity.

In this framework, the dynamics of quantum operators follows the canonical equations of motion \cite{Kowalski2002}. Specifically, any operator $\hat{O}$ evolves in the Heisenberg picture according to:
\begin{equation}
\frac{d \hat{O}}{dt} = \frac{i}{\hbar} [\hat{H}, \hat{O}], \label{Eccanoncero}
\end{equation}
and the evolution of its expectation value $\langle \hat{O} \rangle \equiv {\rm Tr}[\hat{\rho} \hat{O}(t)]$ is given by:
\begin{equation}
\frac{d \langle \hat{O} \rangle}{dt} = \frac{i}{\hbar} \langle [\hat{H}, \hat{O}] \rangle, \label{VM}
\end{equation}
where the average is taken with respect to a suitable quantum density operator $\hat{\rho}$.

On the other hand, the classical conjugate variables $A$ and $P_A$ obey Hamilton's equations, with the generator being the expectation value of the full Hamiltonian:
\begin{subequations}
\label{eqclasgen}
\begin{eqnarray}
\frac{dA}{dt} & = & \frac{\partial \langle \hat{H} \rangle}{\partial P_A}, \label{ds} \\
\frac{dP_A}{dt} & = & -\frac{\partial \langle \hat{H} \rangle}{\partial A} - \eta P_A, \label{clasgenb}
\end{eqnarray}
\end{subequations}
where the parameter $\eta$ is incorporated ad hoc. A positive value of $\eta$ introduces dissipation into the system.

Each operator $\hat{O}_i$ satisfies a commutation relation of the form:
\begin{equation}
\label{conm}
[\hat{H}, \hat{O}_i] = i\hbar \sum_{j=1}^{q} g_{ji}(A, P_A) \, \hat{O}_j, \hskip 1.0cm i = 0, 1, \ldots, q,
\end{equation}
where the coefficients $g_{ji}$ are real numbers. In general, it may be necessary to consider the limit $q \rightarrow \infty$.

In this work, however, we restrict ourselves to a set of operators that form a closed Lie algebra, implying a finite $q$. Consequently, equations (\ref{VM}) and (\ref{eqclasgen}) together define an autonomous system of coupled differential equations—either linear or nonlinear—that provide a consistent dynamical description without violating quantum principles. In particular, the Uncertainty Principle remains preserved at all times. The classical variables $A$ and $P_A$ act as time-dependent parameters for the quantum system. 

Equivalently, this hybrid method preserves the symplectic structure of the classical subsystem because the classical variables $A$ and $P_A$ fully satisfy Hamilton's equations~\eqref{eqclasgen} when $\eta = 0$. As a result, the classical subsystem's time flow remains canonical, while quantum expectation values serve as temporal parameters.
When $\eta > 0$, the classical equations no longer preserve the symplectic structure due to dissipation. This breaks the canonical nature of the phase-space flow, as is typically observed in standard examples of purely classical systems. In this work, we consider an example of such a situation.
 
The initial conditions are specified by an appropriate quantum density operator $\hat{\rho}$, which evolves in time according to the Liouville–von Neumann equation:
\begin{equation}
\label{Liouville}
i\hbar \frac{d \hat{\rho}}{dt} = [\hat{H}(t), \hat{\rho}(t)]~.
\end{equation}

\subsection{Quantum-Classical Hybrid MaxEnt Density Matrix}
\label{Rhogeneral}

We assume (1) complete knowledge of the initial conditions of the classical variables, and  
(2) incomplete knowledge of the quantum system. Under these circumstances, we employ the MaxEnt methodology~\cite{Katz}.  
If the mean values (MV) $\langle \hat{O}_{i} \rangle$ of $q$ operators $\hat{O}_{i}$ are known, the MaxEnt statistical operator $\hat{\rho}$ for our problem takes the form
\begin{equation}
\label{rho}
\hat{\rho} = \exp{\left( -\lambda_{0} \hat{I} - \sum_{i=1}^{q} \lambda_{i} \hat{O}_{i} \right)},
\end{equation}
\noindent
where the $q+1$ Lagrange multipliers $\lambda_{i}$ depend on the classical variables, and $\hat{\rho}$ represents a mixed quantum density matrix.

The Lagrange multipliers are determined so as to satisfy the set of constraints imposed by our prior quantum information  
(i.e., normalization of $\hat{\rho}$ and the $q$ known mean values):
\begin{equation}
\label{constraints}
\langle \hat{O}_{i} \rangle = {\rm Tr} \left[ \hat{\rho} \, \hat{O}_{i} \right], \hskip 1.0cm i = 0, 1, \ldots, q,
\end{equation}
\noindent
where $\hat{O}_{0} = \hat{I}$ is the identity operator. The initial values of the multipliers are obtained by solving the coupled set of equations~\cite{Katz}
\begin{equation}
\label{katze}
\frac{\partial \lambda_0}{\partial \lambda_i} = - \langle \hat{O}_{i} \rangle, \hskip 1.0cm i = 1, 2, \ldots, q,
\end{equation}
which remain valid for all $t$. The expression for $\lambda_{0}$ is
\begin{equation}
\label{landa0}
\lambda_{0} = \ln {\rm Tr} \left[ \exp \left( -\sum_{i=1}^{q} \lambda_{i} \hat{O}_{i} \right) \right].
\end{equation}

The entropy (with $k = 1$) is given by
\begin{equation}
\label{S}
S(\hat{\rho}) = -{\rm Tr} \left[ \hat{\rho} \ln \hat{\rho} \right] = \lambda_{0} + \sum_{i=1}^{q} \lambda_{i} \langle \hat{O}_{i} \rangle,
\end{equation}
\noindent
and is maximized by the MaxEnt statistical operator $\hat{\rho}$~\cite{Katz}.  
If the operators in Eq.~(\ref{rho}) close a Lie algebra, the Lagrange multipliers $\lambda_{j}$ must satisfy the set of differential equations
\begin{equation}
\label{lambda}
\frac{d \lambda_{i}}{dt} = \sum_{j=1}^{q} g_{ij}(A, P_A) \, \lambda_{j}, \hskip 1.0cm i = 0, 1, \ldots, q,
\end{equation}
\noindent
in order to ensure that the MaxEnt {\it form} of $\hat{\rho}$ given in Eq.~(\ref{rho}) remains valid for all $t$, satisfying Eq.~(\ref{Liouville})~\cite{Katz,Levine}.  
The dynamical problem thus reduces to solving the system of equations~(\ref{lambda}), with initial conditions for the multipliers obtained via Eqs.~(\ref{katze}).

\section{Semiclassical Hybrid Hamiltonian}
\label{semihybryhamil}

We consider the following quantum-classical hybrid system:
\begin{equation}
\hat{H}= \dfrac{1}{2}\left(\alpha_{1}\hat{x}^2+
\alpha_{2}\hat{p}^2+ \alpha_{3} \hat{L} +\beta_{1}F(A) \hat{I} +\beta_{2}P_{A}^{2}\hat{I}+\gamma V(A)\hat{x}^{2}\right),
\label{Hpaper}
\end{equation}
where $\hat{x}$ and $\hat{p}$ are quantum operators, while $A$ and $P_{A}$ are classical variables. The operator $\hat{L}= \hat{x}\hat{p}+\hat{p}\hat{x}$ represents the quantum correlation. The functions $F(A)$ and $V(A)$ are classical and assumed to possess continuous second-order partial derivatives (see the discussion following system (\ref{class})).

The term $\gamma V(A)\hat{x}^{2}$ represents the interaction between classical and quantum variables and may introduce non-linearity into the system, as can $F(A)$. In the latter case, however, the non-linearity arises solely from classical origins. Although a more general quadratic interaction involving $\hat{x}^2$, $\hat{p}^2$, and $\hat{L}$ could be considered, we adopt the simplified form $\hat{H}_{cl}^{q}= \gamma V(A)\hat{x}^{2}$ without loss of generality. Linear terms are not included because they can be readily eliminated through coordinate translations.

All coefficients may take different signs or even vanish. We pay particular attention to the coefficients $\alpha_i$, which define the nature of the quantum subsystem. As such, the Hamiltonian can represent simple yet physically relevant systems, such as a harmonic oscillator, a free particle, or a system with an inverse parabolic potential.

The Hamiltonian (\ref{Hpaper}) encompasses various models across fields such as quantum computing, quantum information, condensed matter, and quantum optics. It includes, as particular cases, the chaotic system studied in Ref. \cite{previo} and the dissipative model analyzed in Ref. \cite{disipativo}, which we denote as $\hat{H}_1$ and $\hat{H}_2$, respectively.

The Hamiltonians $\hat{H}_1$ and $\hat{H}_2$ differ significantly in their dynamics. $\hat{H}_1$ exhibits chaotic behavior, while $\hat{H}_2$ displays regular dynamics. This distinction arises because $\hat{H}_1$ describes a quantum harmonic oscillator interacting with a classical free particle (an unbounded classical system), whereas $\hat{H}_2$ involves two harmonic oscillators—one quantum and one classical—due to the inclusion of the term $\omega_{cl} P_{A}^2$. Consequently, the contrasting dynamics are evident in Figures \ref{Fig1} and \ref{dis_case}. Moreover, in the case of $\hat{H}_2$, the dissipative parameter $\eta$ in Eq. (\ref{clasgenb}) is non-zero, resulting in dissipative behavior.

We analyze the time evolution of the set $(\hat{x}^2, \hat{p}^2,\hat{L})$, which forms a closed Lie algebra. This minimal set encodes information about the uncertainty principle via the invariant $I$ (see Eq. (\ref{eqInv})). The equations for their expectation values (EVs) are:
\begin{subequations}
\begin{align}
& \dfrac{d\langle \hat{x}^2\rangle}{dt} =\alpha_{2} \langle \hat{L}\rangle +2\alpha_{3}\langle \hat{x}^2\rangle,\\
& \dfrac{d\langle \hat{p}^2\rangle}{dt}  =-\left(\alpha_{1} +\gamma V(A)\right)\langle \hat{L}\rangle- 2\alpha_{3}\langle\hat{p}^2\rangle,\\
& \dfrac{d\langle \hat{L}\rangle}{dt}  = 2\left(\alpha_{2} \langle\hat{p}^{2}\rangle-\left(\alpha_{1}+\gamma V(A)\right)\langle\hat{x}^{2}\rangle\right),
\end{align}
\label{Eccanon}
\end{subequations}
while the classical variables governed by Eqs. (\ref{eqclasgen}) satisfy:
\begin{subequations}
\begin{align}
&\dfrac{dA}{dt}=\beta_{2} P_{A}, \\
&\dfrac{dP_{A}}{dt}=-\left(\beta_{1}  \frac{dF(A)}{dA} +\gamma \frac{dV(A)}{dA}\langle\hat{x}^{2}\rangle \right) - \eta P_{A}. \label{PA}
\end{align}
\label{eqclasgen1}
\end{subequations}
Equations (\ref{Eccanon}) and (\ref{eqclasgen1}) form an autonomous set of coupled first-order ordinary differential equations (ODEs), generally nonlinear. When $\eta > 0$, the system exhibits dissipative behavior.

We define the invariant $I$ as:
\begin{equation}
I=\langle \hat{x}^{2}\rangle \langle \hat{p}^{2}\rangle -\dfrac{\langle \hat L\rangle ^{2}}{4} \geq \dfrac{\hbar ^{2}}{4},
\label{eqInv}
\end{equation}
which remains conserved under the dynamics of systems (\ref{Eccanon})–(\ref{eqclasgen1}) in both conservative and dissipative regimes (see Ref. \cite{disipativo}). This invariant is related to the uncertainty principle and quantifies deviations from classicality, where the classical condition corresponds to $I = x^2 p^2 - L^2 /4 = 0$ (a trivial invariant).

To study the classical limit of this system, we compare it with the classical analogue of the Hamiltonian (\ref{Hpaper}):
\begin{equation}
H_{cl}=\dfrac{1}{2}\left(\alpha_{1}x^2+
\alpha_{2}p^2+ \alpha_{3} L +\beta_{1} F(A)  +\beta_{2}P_{A}^{2}+\gamma V(A)x^{2}\right).
\label{Hcl}
\end{equation}
The corresponding classical equations, derived from Hamilton's equations with $L=2xp$, are:
\begin{subequations}
\begin{align}
&\dfrac{dx^{2}}{dt} =\alpha_{2}L +2\alpha_{3}x^2,\\ 
&\dfrac{dp^{2}}{dt}=-\left(\alpha_{1}+\gamma V(A)\right)L -2\alpha_{3}p^2, \\
&\dfrac{dL}{dt}=2\left(\alpha_{2}p^{2}-\left(\alpha_{1}+\gamma V(A)\right)x^{2}\right),\\
&\dfrac{dA}{dt}=\beta_{2} P_{A}, \\
&\dfrac{dP_{A}}{dt}=-\left(\beta_{1}  \frac{dF(A)}{dA} +\gamma\frac{dV(A)}{dA} x^{2}\right) - \eta P_{A}.
\end{align}
\label{class}
\end{subequations}
We emphasize that the classical limit is attained when the solutions of systems (\ref{VM}) and (\ref{eqclasgen}) coincide with those of their classical counterpart, namely Eqs. (\ref{Eccanon}) and (\ref{eqclasgen1}) with Eqs. (\ref{class}).

In this work, we analyze the limit $I \rightarrow 0$. A well-known theorem on ODEs ensures continuous dependence of solutions on initial conditions, provided the Lipschitz condition is satisfied \cite{ODE}. If the system exhibits continuous dependence on its variables (hence the requirement for second-order continuity of $F(A)$ and $V(A)$ with respect to the classical variables $A$ and $P_A$) and its solutions remain bounded as $t \rightarrow \infty$, the condition holds. This is also valid for many common unbounded physical systems.

This guarantees the existence of the limit $I \rightarrow 0$ (see Ref. \cite{Kowalski2002}). For further details, refer to Appendix \ref{AppeC}.

\subsection{MaxEnt Density Operator for our Quantum-Classical Hybrid Hamiltonian}
\label{maxentdenshh}

We choose the operators $\hat O_{1} = \hat{x}^{2}$, $\hat O_{2} = \hat{p}^{2}$, and $\hat O_{3} = \hat{L}$.
As mentioned above, this set is the minimal one that includes information regarding the uncertainty principle (through the invariant $I$). We propose a statistical operator equivalent to that of Ref. \cite{previo}. By this, we mean the same functional form, but with Lagrange multipliers whose time evolution depends on the particular Hamiltonian that satisfies Eq. (\ref{Hpaper}).

The normalized MaxEnt statistical operator $\hat \rho$ is expressed in terms of the Lagrange multipliers $\lambda_{i}(t)$ as \cite{previo,caosclasico}
\begin{equation}
\label{rholevel1}
\hat \rho(t) = \exp\left[ -\left(\lambda_{0} \hat
I + \lambda_1 (t) \hat{x}^{2} + \lambda_2 (t) \hat{p}^{2} + \lambda_3  (t)\hat{L} \right)\right],
\end{equation}
for all $t$, since it satisfies the Liouville–von Neumann equation. This is because the set $\hat O_{1} = \hat{x}^{2}$, $\hat O_{2} = \hat{p}^{2}$, $\hat O_{3} = \hat{L}$ forms a Lie algebra (see Refs. \cite{Katz,Levine}). Once Eqs. (\ref{lambda}) have been solved, we obtain the time dependence of any expectation value via Eqs. (\ref{constraints}), that is,
\begin{equation}
\langle \hat O \rangle = {\rm Tr} \left[ \hat \rho (t)\, \hat O \right].
\label{constraints2}
\end{equation}

In the semiclassical case, Eqs. (\ref{lambda}) depend on the classical variables, which requires consideration of the corresponding classical evolution equations (\ref{eqclasgen})–(\ref{eqclasgen1}). The presence of the term $\langle \hat{x}^{2} \rangle$ in the equation for $P_A$ (Eq. \ref{PA}) also introduces a dependence on the mean values, but this situation can be resolved through
\begin{equation}
\label{O4,L}
\langle \hat{x}^{2} \rangle (t)= {\rm Tr} [\hat \rho(t)
\hat{x}^{2} ],
\end{equation}
as we shall see below.

To address the problem at hand, it is convenient to apply an appropriate transformation. We use the unitary transformation (\ref{Trans}) from Appendix \ref{AppeA}, introduced in Eqs. (16) for the Hamiltonian given by Eq. (3) of Ref. \cite{previo}. This transformation has the same functional form, but the time evolution of the Lagrange multipliers now depends on the general Hamiltonian (\ref{Hpaper}). For simplicity, from now on we will use $\lambda_{i}$ instead of $\lambda_{i}(t)$. This transformation allows us to express Eq. (\ref{rholevel1}) in the convenient form \cite{previo}
\begin{equation}
\label{rholevel1I}
\hat \rho(t) = \exp( -\lambda_{0}) \exp \left[-  I_\lambda  \left( \hat{X}^{2} +  \hat{P}^{2} \right) \right].
\end{equation}

For the convergence of Eqs. (\ref{rholevel1}) and (\ref{rholevel1I}), $\lambda_1$, $\lambda_2$, and $\lambda_{1} \lambda_{2} - {\lambda_{3}}^{2}$ must be positive. Then, $\lambda_V$ and $\lambda_T$ are also positive, and
\begin{equation}
I_{\lambda}  =  \sqrt{ \lambda_{1} \lambda_{2} - {\lambda_{3}}^{2} },
\label{Ilambda}
\end{equation}
is well defined \cite{previo}.

Transformation (\ref{Trans}) preserves the commutation relations. Thus, $I$ is also preserved.

We can now rewrite the dynamical equations for the multipliers using Eq. (\ref{rel.mv4,xcuad}), as a closed and autonomous system of equations involving both the multipliers and the classical variables:
\begin{subequations}
\label{lamb+s,p}
\begin{eqnarray}
\frac{d\lambda_{1}}{dt} & = &2\left(\alpha_{1}+\gamma V(A)\right)\lambda_{3}-2\alpha_{3}\lambda_{1},\\
\frac{d\lambda_{2}}{dt} & = & 2\left(\alpha_{3}\lambda_{2}-\alpha_{2}\lambda_{3}\right),\\
\frac{d\lambda_{3}}{dt} & = & \left(\alpha_{1}+\gamma V(A)\right) \lambda_{2}-\alpha_{2}\lambda_{1},\\
\frac{dA}{dt} & = & \beta_{2}P_{A}, \\
\frac{dP_{A}}{dt} & = &-\left(\beta_{1}  \frac{dF(A)}{dA} +\gamma \frac{dV(A)}{dA} \frac{T(I_{\lambda})}{I_{\lambda}}\lambda_{2}\right) - \eta P_{A}, \label{lamb+s,p,p}
\end{eqnarray}
\end{subequations}
where $T(I_{\lambda})$ is given by Eq. (\ref{tilambda}). It follows from Eq. (\ref{lamb+s,p}) that the quantity $I_{\lambda}$ is a constant of motion, i.e., $dI_{\lambda} / dt = 0$. The square root of this invariant plays the role of $I$, expressed in terms of the $\lambda$'s. The temporal evolution in Eq. (\ref{rholevel1I}) emerges through the new operators $\hat X$ and $\hat P$, via the $\lambda$'s that appear as coefficients in their definition and contain all the relevant information regarding the classical degrees of freedom. Solving Eq. (\ref{lamb+s,p}) allows us to determine the time dependence of any expectation value via Eq. (\ref{constraints2}).


\section{Classical Limit. Previous results}
\label{previousres}

In several works, we studied the classical limit via the Eqs. (\ref{VM}) and (\ref{eqclasgen})  as a function of the relative energy $E_r$, defined as $ E_{r}=\frac{|E|}{I^{1/2}} $, where
$E=\langle H\rangle$ is the total energy of the system. For example, in Refs. \cite{Kowalski2002,BP,energia}.  The classical limit is obtained for $E_r \rightarrow \infty$. In practice, this means $|E| \gg I^{1/2} $. 
This analysis was done for the Hamiltonian $\hat{H_1}~=~\frac{1}{2}\left(~\frac{\hat{p}^{2}}{m_{q}}~+~
\frac{{P_{A}}^{2}}{m_{cl}}~+~ m_{q}\omega ^{2}\hat{x}^{2}~\right)$, particular case of (\ref{Hpaper}).  Here $\omega^2 = \omega_q^2 + e^{2} A^{2}$. $\hat{H_1}$ studied by Cooper et al. \cite{cooper94}, represents the zero-momentum part of the problem of pair production of charged mesons by a strong external electric field. In Ref. \cite{previo}, we investigated numerically the dynamics described by a density operator of the form (\ref{rholevel1}).

\begin{figure}
\centering
\subfloat[]{\includegraphics[width=5.1cm]{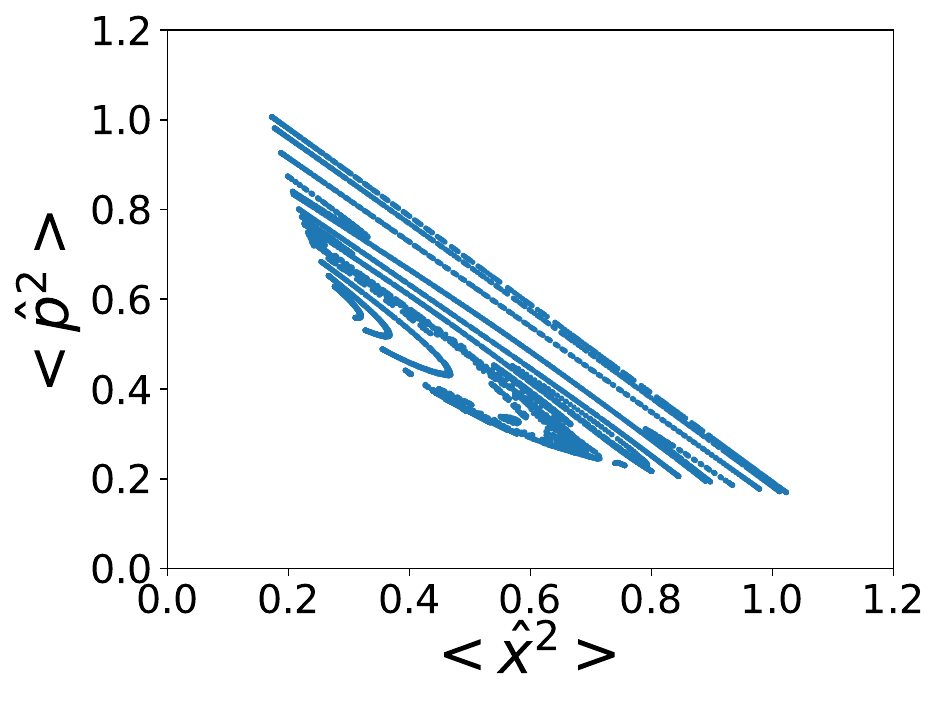}}
\subfloat[]{\includegraphics[width=5.1cm]{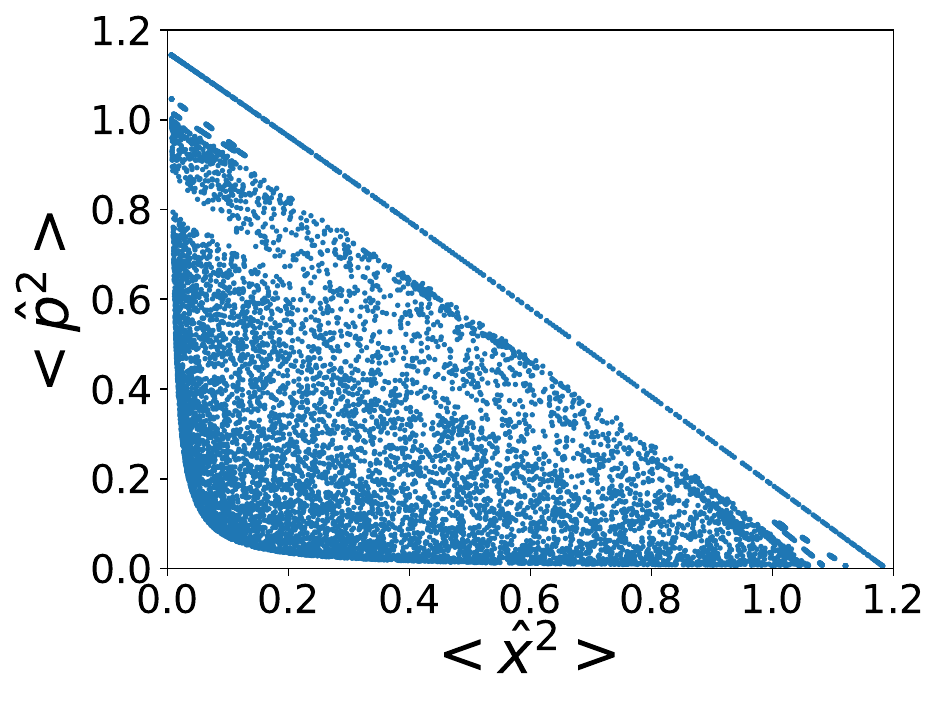}}
\subfloat[]{\includegraphics[width=5.1cm]{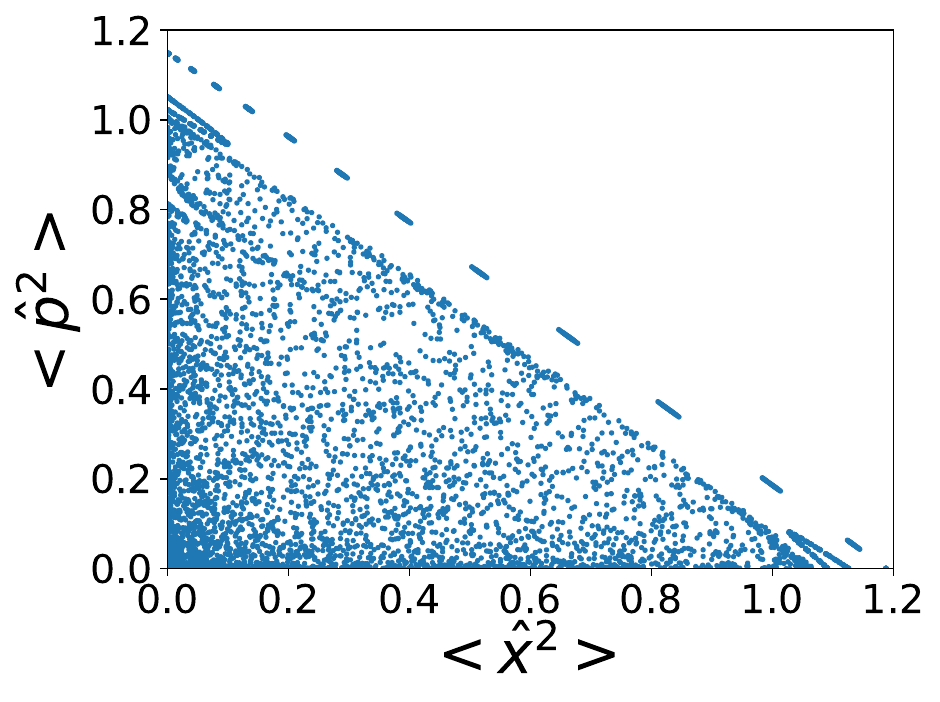}}

\subfloat[]{\includegraphics[width=5.2cm]{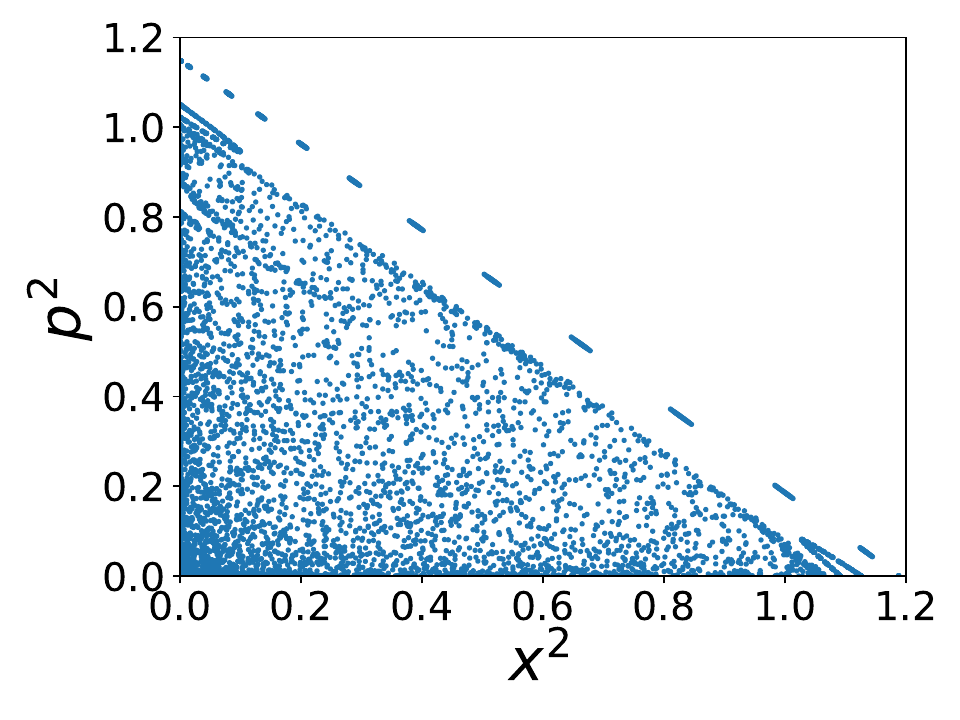} }
\captionsetup{justification=justified, singlelinecheck=false} 
\caption{Poincaré sections for the case $A = 0$, corresponding to the specific Hamiltonian 
$\hat{H}_1 = \frac{1}{2}\left(\frac{\hat{p}^{2}}{m_q} + \frac{P_A^{2}}{m_{cl}} + m_q \omega^2 \hat{x}^{2}\right)$ from Eq.~(\ref{Hpaper}), where $\omega^2 = \omega_q^2 + e^2 A^2$. The energy is set to $E = 0.6$.
Figures (a) and (b) correspond to $I = 0.17361$ and $I = 2.341311 \times 10^{-5}$, respectively, both within the transition (mesoscopic) regime. These sections are bounded by the curves 
$\frac{\langle \hat{p}^{2} \rangle}{m_q} + m_q \omega_q^2 \langle \hat{x}^{2} \rangle = 2E$, 
associated with the total energy, and 
$\langle \hat{x}^2 \rangle \langle \hat{p}^2 \rangle = I$, which reflects the Uncertainty Principle (see Eq.~(\ref{eqInv})). These plots demonstrate the coexistence of chaotic dynamics with the Heisenberg Principle, illustrating semiclassical chaos.
Figure (c) corresponds to $I = 3.6 \times 10^{-23}$.
Figure (d) shows $x^2$ versus $p^2$, representing the limits 
$\lim\limits_{\hbar \rightarrow 0} \lim\limits_{I \rightarrow \hbar^2/4} \langle \hat{x}^2 \rangle$ and 
$\lim\limits_{\hbar \rightarrow 0} \lim\limits_{I \rightarrow \hbar^2/4} \langle \hat{p}^2 \rangle$ 
(see Section \ref{currentres}), which coincide with the classical case $I = 0$. This plot is bounded solely by 
$\frac{p^2}{m_q} + m_q \omega_q^2 x^2 = 2E$, and illustrates classical chaos.
All results were validated by confirming the time invariance of the dynamical quantities $E$ and $I$, with a numerical precision of $10^{-80}$.}
\label{Fig1}
\end{figure}

\begin{figure}
\centering
\subfloat[]{\includegraphics[width=5.1cm]{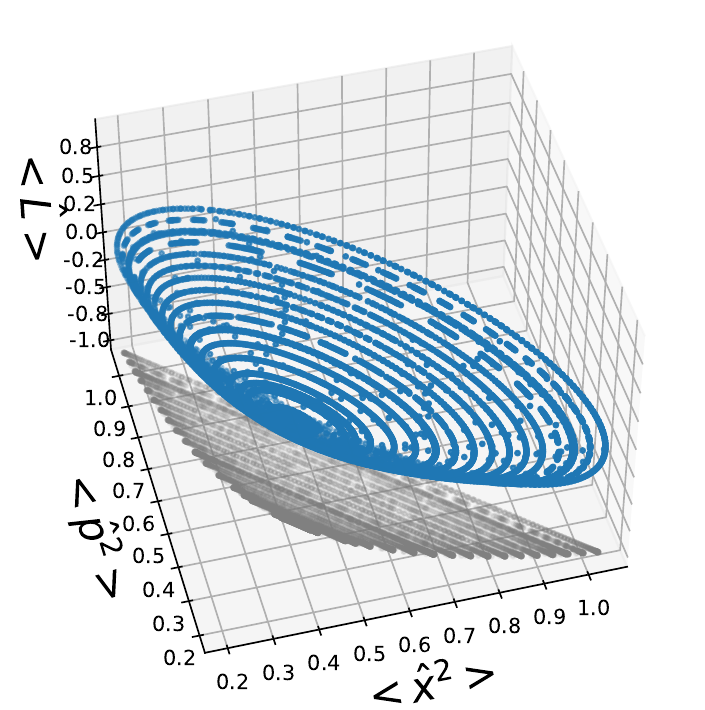}}
\subfloat[]{\includegraphics[width=5.1cm]{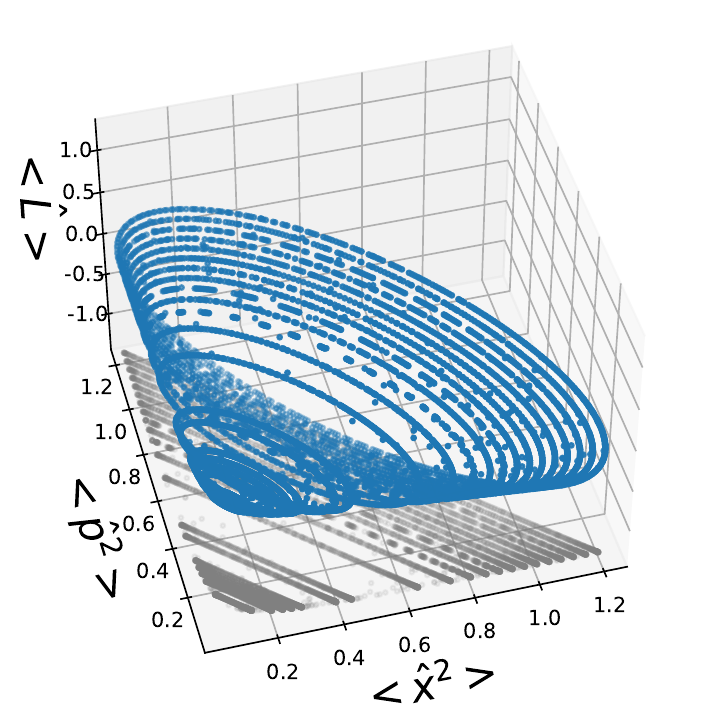}}
\subfloat[]{\includegraphics[width=5.1cm]{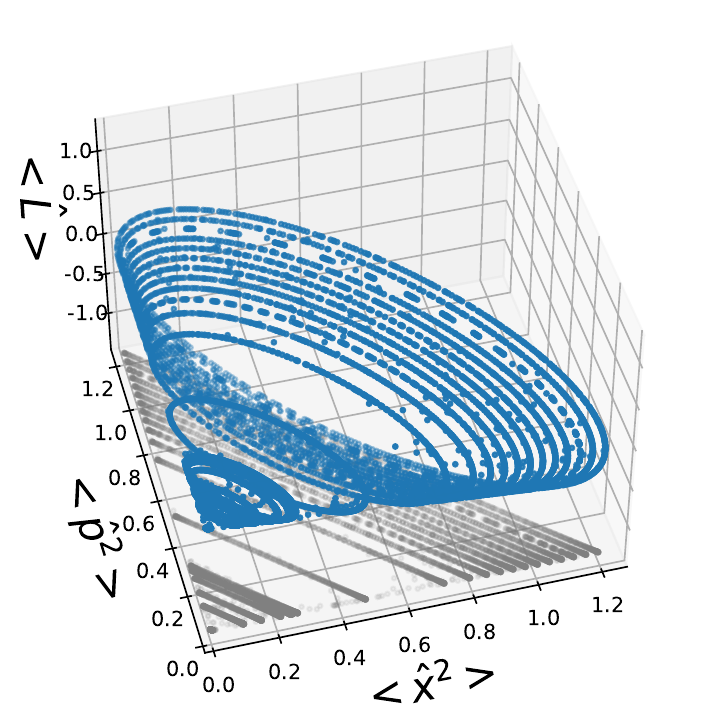}}

\subfloat[]{\includegraphics[width=5.2cm]{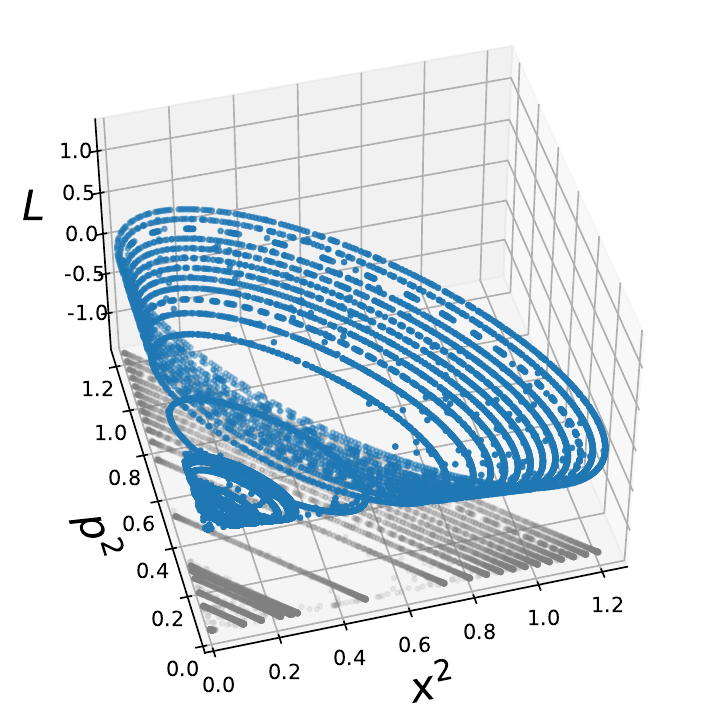}}
\caption{3D Poincaré sections with $A = 0$ and their projections, corresponding to the specific case $\hat{H}_2 = \frac{1}{2} \left( \omega_q (\hat{x}^2 + \hat{p}^2) + \omega_{cl}(A^2 + P_A^2)\hat{I} + e_q^{cl} A^2 \hat{x}^2 \right)$ from Eq.~(\ref{Hpaper}). This system is highly relevant in condensed matter physics. We present an example in the dissipative regime, where energy is not conserved. Consequently, only one dynamical invariant remains, enabling the construction of 3D Poincaré sections. The initial energy is $E(0) = 0.6$. Figures (a), (b), and (c) display the Poincaré sections in the space of $\langle \hat{x}^2 \rangle$, $\langle \hat{p}^2 \rangle$, and $\langle \hat{L} \rangle$, along with their respective projections onto the $\langle \hat{x}^2 \rangle$–$\langle \hat{p}^2 \rangle$ plane (see details in Ref.~\cite{disipativo}). The corresponding values of $I$ are $I = 0.17361$, $2.34131 \times 10^{-5}$, and $3.6 \times 10^{-23}$, respectively. Figure (d) corresponds to $I = 0$, which represents the classical limit: $\lim\limits_{\hbar \rightarrow 0} \lim\limits_{I \rightarrow \hbar^2/4} \langle \hat{x}^2 \rangle$, $\lim\limits_{\hbar \rightarrow 0} \lim\limits_{I \rightarrow \hbar^2/4} \langle \hat{p}^2 \rangle$, and $\lim\limits_{\hbar \rightarrow 0} \lim\limits_{I \rightarrow \hbar^2/4} \langle \hat{L} \rangle$. The dissipative parameter is $\eta = 0.05$. Note the curve $\langle \hat{x}^2 \rangle \langle \hat{p}^2 \rangle = I$ (see Eq.~(\ref{eqInv})) in the projections of Figs.~(a), (b), and (c).}
\label{dis_case}
\end{figure}

 We have shown the existence of the limit, by increasing $E_r$ (for example, decreasing $I$ \cite{Kowalski2002,BP} or increasing $E$ \cite{energia}). In this way, the system passes through three zones: a quasiquantal one, a transitional (or mesoscopic),  and a classical region (see Figs. \ref{Fig1} and \ref{dis_case} of Refs. \cite{Kowalski2002,BP,previo}). As $E_r$ grows, chaos emerges in the transitional sector, which increases significantly in the classic one. 

In Ref.\cite{disipativo} we have considered the conservative and dissipative regimes corresponding to the Hamiltonian  
$\hat{H}_2 = \frac{1}{2} \left( \omega_{q}(\hat{x}^2 + \hat{p}^2) + \omega_{cl}(A^2 + P_A^2) \hat{I} + e_{q}^{cl} A^2 \hat{x}^2 \right)$,  
also a particular case of (\ref{Hpaper}). System of great importance in condensed matter. This analysis was performed using Eqs. (\ref{VM}) and (\ref{eqclasgen}). We find the existence of the three zones as a function of $E_r$, as in the previous case, both in the conservative and dissipative regimes. In the dissipative case, we replace $E$ by $E(0)$. Because energy is not conserved, we need 3D Poincar\'e sections (see details in Ref. \cite{disipativo}). Figs. \ref{dis_case} show 3D Poincar\'e sections and their respective projections for the dissipative case. This kind of Hamiltonian is commonly used in the interaction of matter with electromagnetic fields \cite{Milonni, meystre07}.

The calculations for Figs. \ref{Fig1}(a-b-c) and Figs. \ref{dis_case}(a-b-c) were made using the mean values from Eqs. (\ref{VM}) and (\ref{eqclasgen}), and coincide with the results obtained via the Lagrange system of equations (\ref{lambda}) together with (\ref{constraints}) (see below). Figs. \ref{Fig1}(d) and \ref{dis_case}(d) represent the fully classical solutions of the analogous classical Hamiltonian of (\ref{Hpaper}).

\begin{figure}
\centering
\includegraphics[width=7cm]{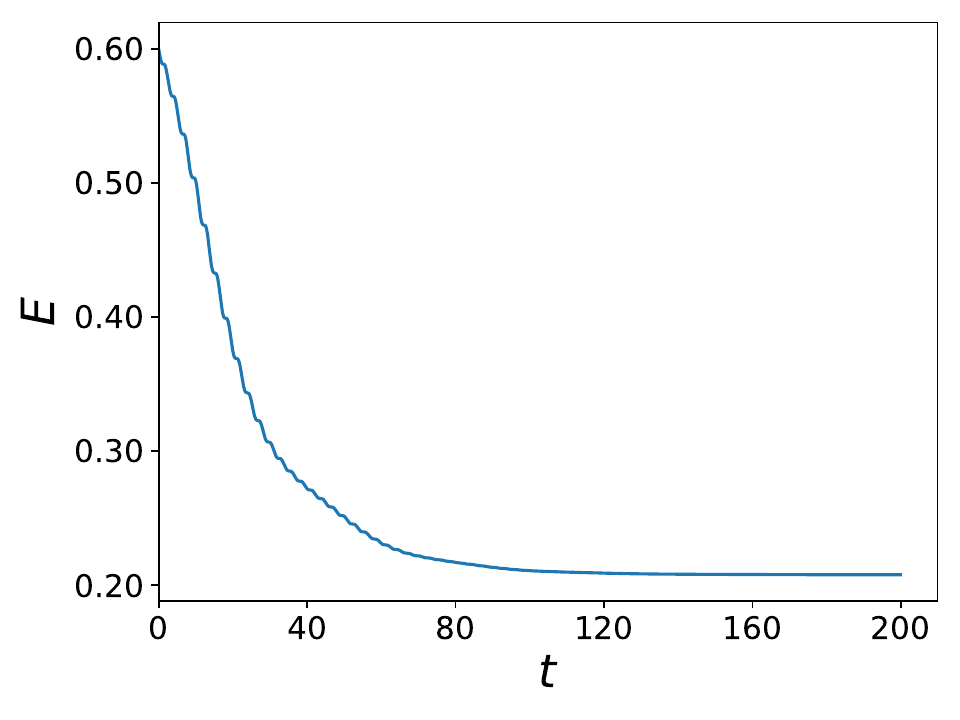}
\caption{Energy as a function of time in the dissipative regime of the Hamiltonian shown in Fig. \ref{dis_case}, for the specific initial condition $E(0) = 0.6$. The final energy remains non-zero, consistent with the Heisenberg Uncertainty Principle \cite{disipativo}. } \label{Hvst}
\end{figure}

In Fig. \ref{Hvst} we plot energy vs. time. It is observed that the final energy is different from zero, which agrees with the Heisenberg uncertainty principle \cite{disipativo}.

\section{Current Results on the Classical Limit}
\label{currentres}

In Ref.~\cite{previo}, we numerically investigated the approach to the classical limit of the Hamiltonian $\hat{H}_1$ using a MaxEnt density operator. In Ref.~\cite{caosclasico}, the classical limit is analyzed analytically. In the present article, we focus on the general Hamiltonian (\ref{Hpaper}) and examine it analytically, aiming to generalize the previous results.

When analyzing the classical limit through the time evolution of the mean values, it suffices to take the limit $I \rightarrow 0$, since relation (\ref{eqInv}) ensures that $\hbar \rightarrow 0$. This is possible because the mean values depend on Planck's constant only through the initial conditions. In contrast, for the density operator, $\hbar$ appears explicitly, and thus we must consider two quantities.

It is convenient to express the relationships between the mean values and the Lagrange multipliers. Using Eqs.~(\ref{rel.mv,L}) and (\ref{I, T(Ilamba)}), we obtain:
\begin{subequations}
\label{lambavlormedio}
\begin{eqnarray}
 I_{\lambda} \, \langle \hat{x}^{2} \rangle &= \sqrt{I}\, \lambda_{2}, \\
 I_{\lambda} \, \langle \hat{p}^{2} \rangle &= \sqrt{I}\, \lambda_{1}, \\
 I_{\lambda} \, \langle \hat{L} \rangle &= -2 \sqrt{I}\, \lambda_{3}.
\end{eqnarray}
\end{subequations}

By combining Eqs.~(\ref{lambavlormedio}) and (\ref{landa0parti}), we obtain the following important relation:
\begin{equation}
\label{I-Ilamba}
I_{\lambda} = \frac{1}{2\,\hbar} \ln \left( \frac{\sqrt{I} + \frac{\hbar}{2}}{\sqrt{I} - \frac{\hbar}{2}} \right),
\end{equation}
which relates $I_{\lambda}$ to $I$. These results were also derived in Ref.~\cite{previo}, and are valid for any density matrix of the form (\ref{rholevel1}).

We numerically analyze the time evolution of the multipliers $(\lambda_{1}, \lambda_{2}, \lambda_{3})$ in the examples provided, with the aim of confirming—via Eqs.~(\ref{lambavlormedio})—the results obtained using the density matrix in the classical limit.

\begin{figure}
\subfloat[]{\includegraphics[width=5cm]{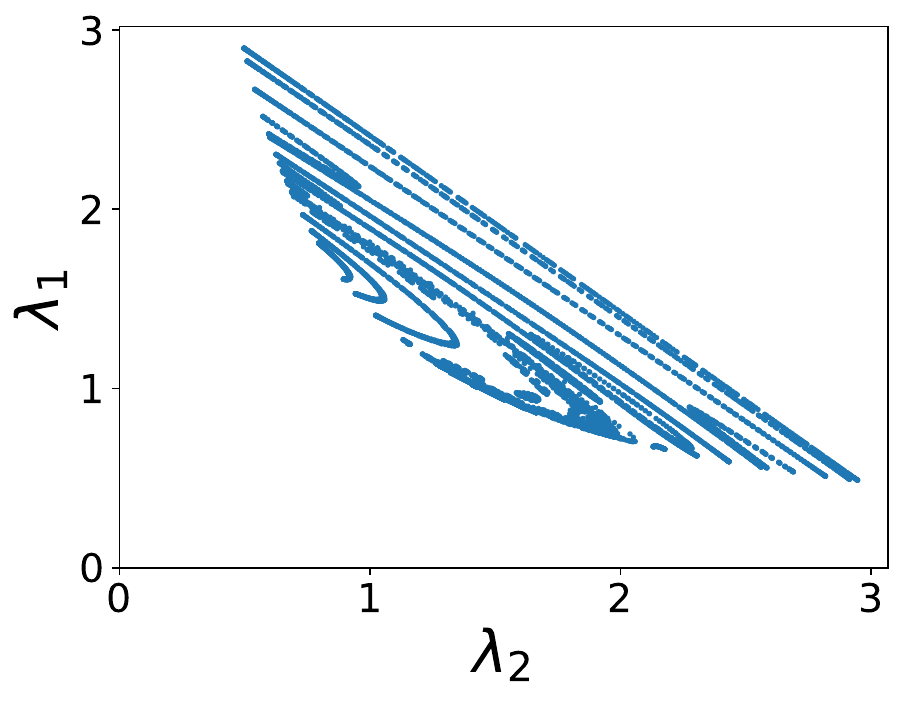}}
\subfloat[]{\includegraphics[width=5.2cm]{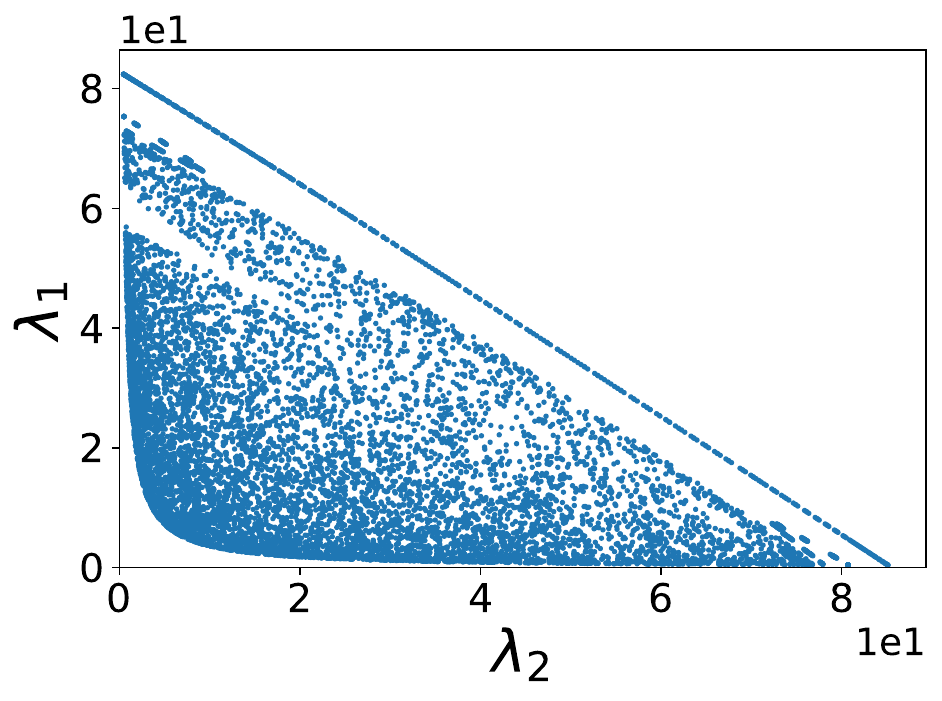}}
\subfloat[]{\includegraphics[width=5.1cm]{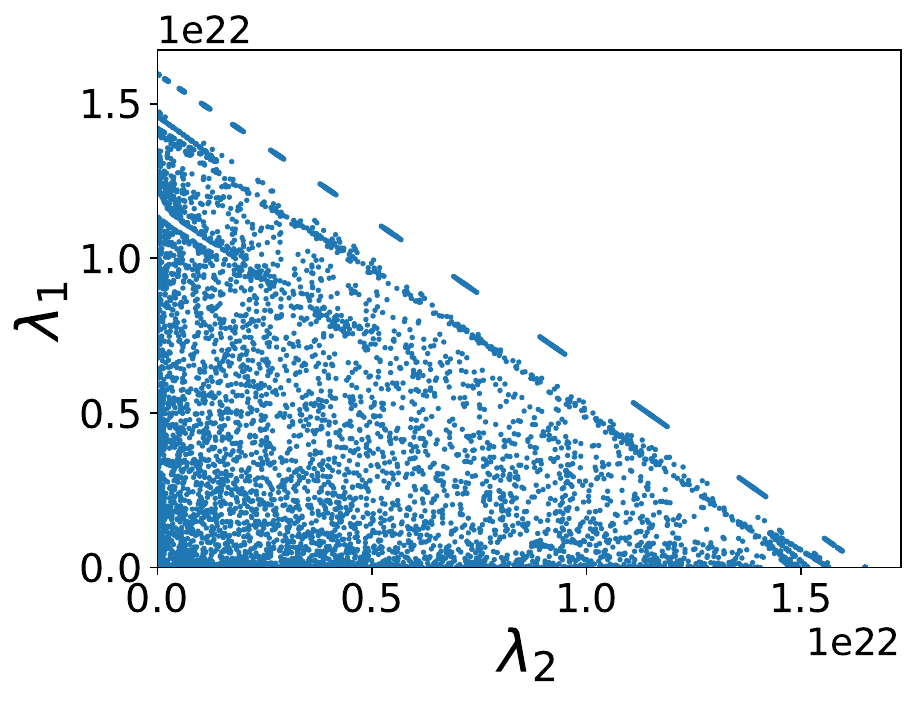}}
\caption{Poincaré sections for $A = 0$, corresponding to the Lagrange multipliers Eqs.~(\ref{lamb+s,p}), in the particular case $\hat{H}_1 = \frac{1}{2}\left(\frac{\hat{p}^{2}}{m_q} + \frac{P_A^{2}}{m_{cl}} + m_q \omega^2 \hat{x}^{2}\right)$. We take $\hbar = 1.0 \times 10^{-40}$ and fix the energy at $E = 0.6$. Fig. (a) corresponds to $I = 0.17361$, Fig. (b) to $I = 2.341311 \times 10^{-5}$, and Fig. (c) to $I = 3.6 \times 10^{-23}$.}  \label{figlambda_c}
\end{figure}

\begin{figure*}
\subfloat[]{\includegraphics[width=5.1cm]{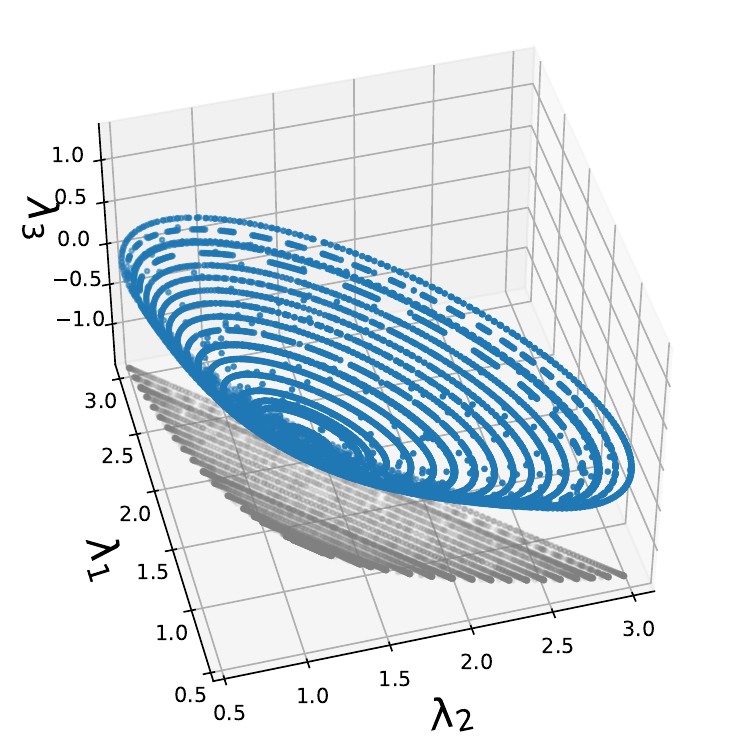}}
\subfloat[]{\includegraphics[width=5.1cm]{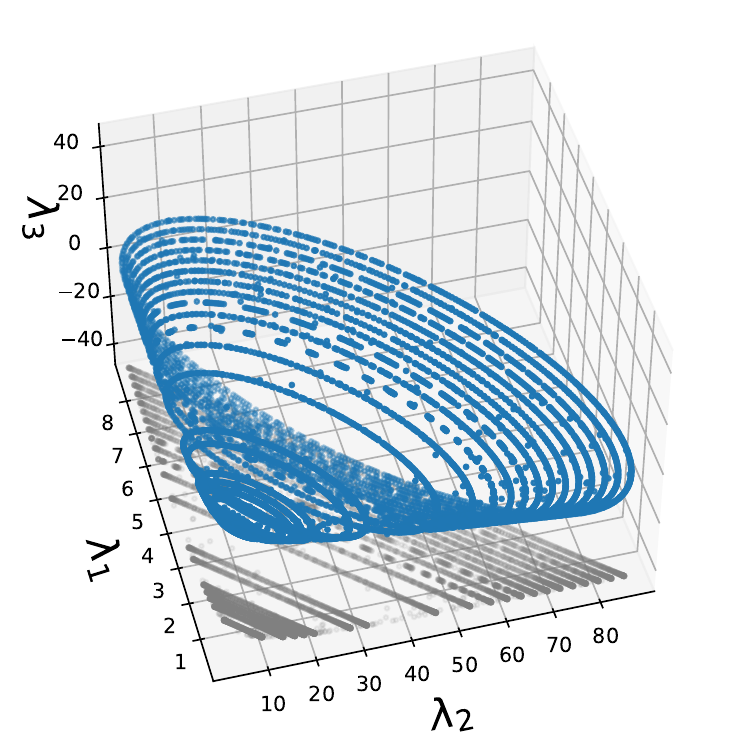}}
\subfloat[]{\includegraphics[width=5.1cm]{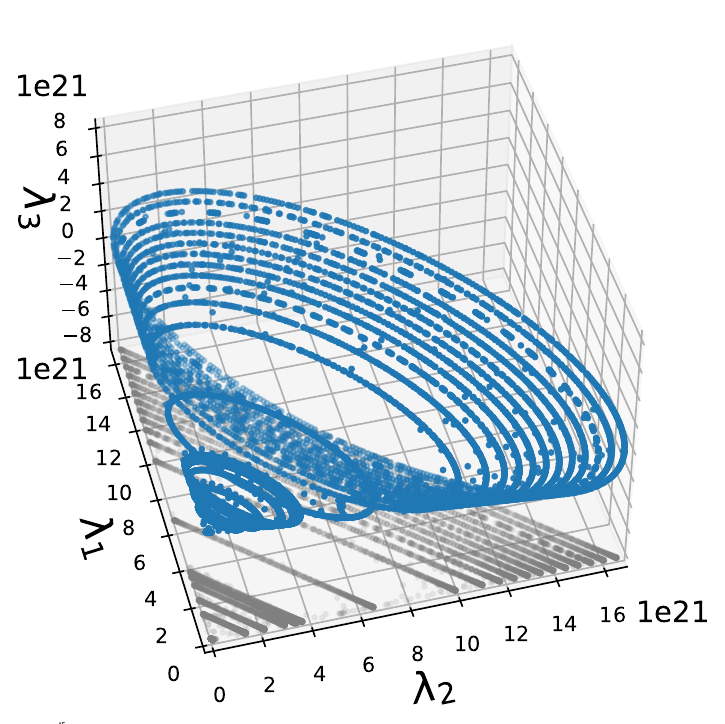}}
\caption{3D Poincaré sections with $A = 0$ and their projections, calculated using Eqs.~(\ref{lamb+s,p}). The Hamiltonian considered is $\hat{H}_2 = \frac{1}{2} \left(\omega_q(\hat{x}^2 + \hat{p}^2) + \omega_{cl}(A^2 + P_A^2)\hat{I} + e_q^{cl} A^2 \hat{x}^2\right)$. We take $\hbar = 1.0 \times 10^{-40}$ and set the initial energy value to $E(0) = 0.6$. This example corresponds to the dissipative case with $\eta = 0.05$. Figures (a), (b), and (c) show the Poincaré sections in the space of $\lambda_1$, $\lambda_2$, and $\lambda_3$, along with their respective projections onto the $\lambda_1$–$\lambda_2$ plane. The corresponding values of $I$ are $0.17361$, $2.34131 \times 10^{-5}$, and $3.6 \times 10^{-23}$, respectively.}
\label{figlambda_d}
\end{figure*}

In Fig. \ref{figlambda_c}, we plot Poincaré sections with $A = 0$, corresponding to the Lagrange multipliers, for the specific case of $\hat{H}_1$. In Fig. \ref{figlambda_d}, we present 3D Poincaré sections with $A = 0$, corresponding to the Lagrange multipliers and their projections, for the case of $\hat{H}_2$. Both figures result from solving Eqs.~(\ref{lamb+s,p}), with $\hbar = 1.0 \times 10^{-40}$.

It is noteworthy that in both figures, the Poincaré sections of the multipliers $(\lambda_{2}, \lambda_{1}, \lambda_{3})$ are identical to those of the mean values $(\langle \hat{x}^{2} \rangle, \langle \hat{p}^{2} \rangle, \langle \hat{L} \rangle)$—with the exchange of axes $(\lambda_{1}, \lambda_{2})$—except for a rescaling of the multipliers, which increases significantly as $I$ grows.

On the other hand, Figs. \ref{Fig1}(a–c) and Figs. \ref{dis_case}(a–c), obtained directly from the mean values $(\langle \hat{x}^{2} \rangle, \langle \hat{p}^{2} \rangle, \langle \hat{L} \rangle)$ via Eqs.~(\ref{Eccanon}) and (\ref{eqclasgen1}), are also derived from the Lagrange system of equations (\ref{lamb+s,p}) and the use of Eqs.~(\ref{lambavlormedio}).

It should be noted that Figs. \ref{Fig1}(d) and \ref{dis_case}(d) cannot be obtained directly from the equations for the multipliers, except in the limit $I \rightarrow 0$.

\begin{figure*}
\centering
\subfloat[]{\includegraphics[width=5.1cm]{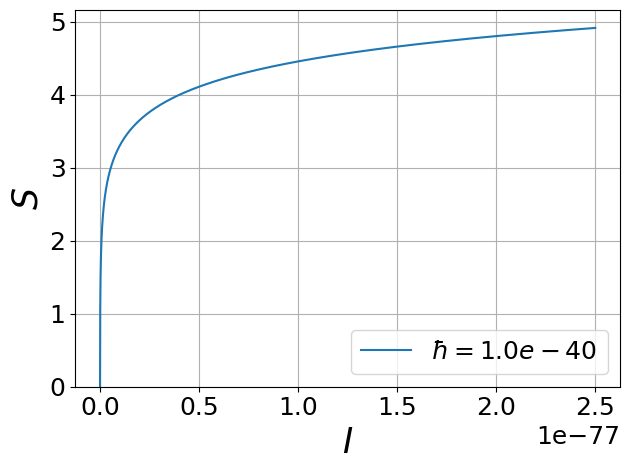}}
\subfloat[]{\includegraphics[width=5.1cm]{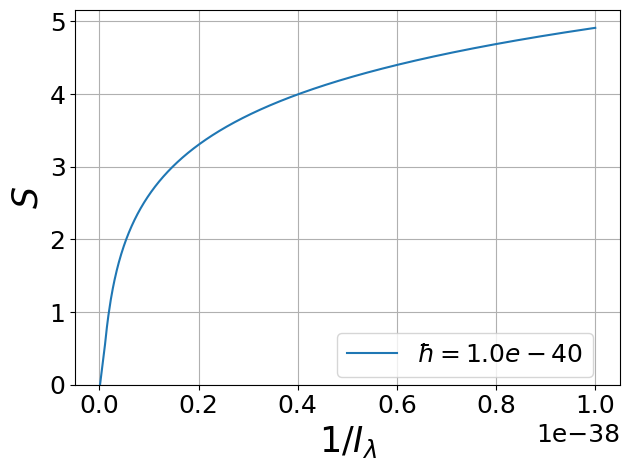}}
\subfloat[]{\includegraphics[width=5.1cm]{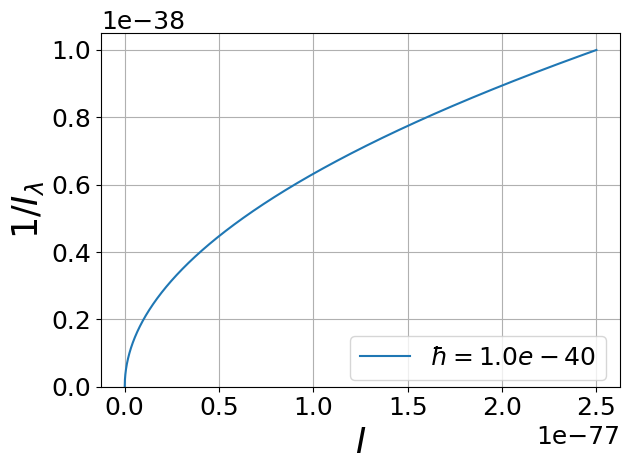}}
\caption{(a) Entropy $S$ as a function of the motion invariant $I$ for $\hbar = 1.0 \times 10^{-40}$. The plot shows that $S$ increases monotonically with $I$. (b) Entropy $S$ versus the pseudo-temperature $1/I_{\lambda}$ for the same value of $\hbar$, showing a monotonic increase of $S$ with pseudo-temperature. (c) Pseudo-temperature $1/I_{\lambda}$ as a function of the motion invariant $I$, also for $\hbar = 1.0 \times 10^{-40}$. It is observed that $1/I_{\lambda} \to 0$ as $I \to 0$.}
\label{SvsI_Ilamb}
\end{figure*}

In going to this limit, we must always respect the inequality (\ref{eqInv}), a restriction that entails some difficulties, so that we face only two possible paths \cite{caosclasico}.

1) The natural a priori solution is to first take $\hbar \rightarrow 0$ (and then $I \rightarrow 0$). Some difficulties are encountered in such instance. Classical statistics and quantum statistics are both compatible with (\ref{eqInv}) for any $\hbar > 0$ (additionally, in the classical case it may be $\hbar = 0$). Therefore, by taking the limit $\hbar \rightarrow 0$, we do not eliminate the statistical aspect of our problem, but only convert quantum statistics into classical statistics. 

2) If we first take a limit on $I$, the direct and natural path is $I \rightarrow \hbar^{2}/4$, that is, bringing $I$ to its lowest possible quantum value (the limit is necessary due to equation (\ref{I-Ilamba})). Then we let $\hbar \rightarrow 0$. This path obviously leads to both quantities tending to zero.

This second option is correct. A numerical verification is achieved by the fact that Fig. \ref{Fig1}(c) and Fig. \ref{dis_case}(c) are obtained from the system of equations (\ref{lamb+s,p}) for the Lagrange multipliers and the use of equations (\ref{lambavlormedio}), even if the value of $\hbar$ is small enough to ``numerically represent" the limit $\hbar \rightarrow 0$.

According to (\ref{landa0parti}), (\ref{lambavlormedio}) and (\ref{I-Ilamba}), we have
\begin{subequations}
\label{infty}
\begin{eqnarray}
\lim\limits_{\hbar \rightarrow 0}\, (\lim\limits_{I \rightarrow \hbar^{2}/4} I_{\lambda}) &=& \infty, \\
\lim\limits_{\hbar \rightarrow 0}\, (\lim\limits_{I \rightarrow \hbar^{2}/4} \hbar \, I_{\lambda}) &=& \infty, \label{infty4} \\
\lim\limits_{\hbar \rightarrow 0}\, (\lim\limits_{I \rightarrow \hbar^{2}/4} \lambda_i) &=& \infty, \quad i = 0, 1, 2, \label{infty1} \\
\lim\limits_{\hbar \rightarrow 0}\, (\lim\limits_{I \rightarrow \hbar^{2}/4} | \lambda_3 |) &=& \infty.
\end{eqnarray}
\end{subequations}
Logically, these are the same relations as those obtained in Ref. \cite{caosclasico}, since the Hamiltonian $\hat{H}_1$ is a particular case of (\ref{Hpaper}). The same applies to the Hamiltonian $\hat{H}_2$ (result obtained for the first time) and to all Hamiltonians represented by (\ref{Hpaper}).

Note that when $I$ tends to its minimum possible value $\hbar^{2}/4$, $\hat{\rho}$ (Eq. \ref{rholevel1I}) tends to its ground state. Thus, considering the \textit{pseudo-temperature} $1/I_{\lambda}$, we ascertain that $1/I_{\lambda} \rightarrow 0$. We remark that $I_{\lambda}$ depends on both the classical variables and the initial conditions for the EVs. These results also hold for $\hbar \rightarrow 0$. 

We see from Eq. (\ref{infty1}) that $\exp(-\lambda_0)$ tends to $\infty$. However, a closer scrutiny of the asymptotic behavior of $\lambda_0$ in Eq. (\ref{landa0parti}) ascertains that $\exp(-\lambda_0) \sim \exp(\hbar \, I_{\lambda})$. Thus, the eigenvalues (\ref{autovalores}) tend to asymptotic values of the form $\exp(-2 n \hbar I_{\lambda})$, $n = 0, 1, 2, \ldots$.  

Keeping in mind Eq. (\ref{infty4}), we find the classical limit of $\hat{\rho}$ given by Eq. (\ref{rholevel1I}). Thus, at the classical limit, $\hat{\rho}$ (in its eigen-basis) has the associated density matrix $\mathcal{R}(t)$.

\begin{equation}
\mathcal{R}(t)=\left(\begin{array}{cc}
     1 \,\,\, 0 \,\,\, 0 \,\,\,\ldots \\
     0 \,\,\, 0 \,\,\, 0 \,\,\,\ldots \\
     0 \,\,\, 0 \,\,\, 0 \,\,\,\ldots \\
     \mathbf{\vdots}
\end{array}\right).
\label{Ademo}
\end{equation}

We have found that in the mixed density matrix in the process toward the limit $I \rightarrow 0$, decoherence occurs. In particular, in this limit, a pure matrix (\ref{Ademo}) of a single state (the fundamental state) is obtained. That is, a measurement occurs. This state will represent the fully classical time evolution.

The expectation values $ \langle \hat {X}^{n} \hat{P}^{m} \rangle $ will be null at all times, which is of a trivial classic nature. Furthermore, the EVs of the set ($\hat{x}^{2}$, $\hat{p}^{2}$, $\hat{L}$) will evolve asymptotically with the classical equations corresponding to the classical counterpart of our quantum Hamiltonian (\ref{Hpaper}).
Any other asymptotic value of a given MV can be calculated using equations (\ref{constraints2}) and ($\ref{Trans}$) and (\ref{Ademo}).

As a proof of the correctness of the results, it is easy to see that $I$  calculated with $\rho(t)$ given by (\ref{Ademo}) vanishes. Denoting the ground state by $|0>$, we have $ <0|\hat X^2|0>=<0|\hat P^2|0>= \lim\limits_{\hbar \rightarrow 0} \hbar/2$ and $<0|\hat L|0>=0$, so that $I=0$.
 Moreover,  via the MaxEnt expression $S(\hat \rho) =-{\rm Tr}\;[\; \hat \rho  \;  \ln \hat \rho \;] \; = \lambda_{0}
 + \sum_{i=1}^{3}\lambda_{i} \langle \hat O_{i} \rangle$, we obtain for the entropy
 \begin{equation}
   S=  \lambda_0 + 2 \, I_{\lambda} \sqrt{I},
  \end{equation}
 which is an increasing and concave function of $I$, with asymptotic value $S=0$ when $I \rightarrow 0$, as expected for a pure state. In Figure \ref{SvsI_Ilamb}(a), we observe this behavior for the value of $\hbar=1.0\times 10^{-40}$. 

 In this way, the density operator gradually becomes less and less mixed, as $I$ tends to zero, until it is represented by a pure-state density matrix.

 Figure \ref{SvsI_Ilamb}(b) also shows that the entropy $S$ as a function of the pseudo-temperature $1/I_{\lambda}$ increases monotonically and is concave.  
 
 Finally, Figure \ref{SvsI_Ilamb}(c) shows in an equivalent way how the pseudo-temperature $1/I_{\lambda}$, is also an increasing and concave function of $I$. It tends to zero when $I$ decreases.

\section{Conclusions}
\label{Concl}

In this article, we analytically examined the classical limit of a broad class of non-linear quantum-classical hybrid systems, composed of a classical subsystem interacting with a quantum one. The fundamental requirements for such systems are:(i) the quantum operators involved must satisfy a Lie algebra, and  (ii) the classical terms in the Hamiltonian (\ref{Hpaper}) must possess continuous second partial derivatives with respect to the classical variables $A$ and $P_A$.

We employed a MaxEnt density operator under conditions of {\it incomplete} quantum a priori information, extending the framework developed in Ref. \cite{caosclasico}.

Previous studies explored the classical limit of quantum systems using Eqs. (\ref{VM}) and (\ref{eqclasgen}), analyzing the behavior in terms of the motion invariant $I$—a quantity linked to the Heisenberg principle \cite{Kowalski2002,BP,energia,previo,caosclasico,disipativo}—and the total energy $E$ \cite{energia}. This led to the identification of three distinct regimes along the path to classicality, characterized by the parameter $E_{r} = \frac{|E|}{I^{1/2}}$: a quasi-classical region, a transitional (mesoscopic) region, and a fully classical zone. Section \ref{previousres} presents two numerical examples illustrating this transition for different Hamiltonians.

In the present work, we focused on the analytical treatment of the special case $E_r \rightarrow \infty$, achieved by keeping $E$ fixed while decreasing $I$.

We analyzed the normalized MaxEnt statistical operator $\hat{\rho}$, expressed in terms of time-dependent Lagrange multipliers $\lambda_{i}(t)$ (\ref{rholevel1}), which satisfies the Liouville–von Neumann equation.

Our approach, consistent with our objectives, involves first taking the limit $\lim I \rightarrow \hbar^{2}/4$, the minimum value allowed by quantum mechanics, followed by the limit $\hbar \rightarrow 0$.

Numerical verification is provided by Figs. \ref{Fig1}(c) and \ref{dis_case}(c), which were obtained from the system of equations (\ref{lamb+s,p}) for the Lagrange multipliers and the expectation value equations (\ref{lambavlormedio}), even for values of $\hbar$ small enough to numerically approximate the limit $\hbar \rightarrow 0$.

We found that, as $I \rightarrow 0$, decoherence naturally emerges in the mixed density matrix. In this limit, the system evolves toward a pure state matrix (\ref{Ademo}) representing the fundamental state—effectively a measurement—which governs the classical time evolution.

The expectation values of the observables ($\hat{x}^{2}$, $\hat{p}^{2}$, $\hat{L}$) asymptotically follow the classical equations derived from the classical analogue (\ref{Hcl}) of the quantum-classical hybrid Hamiltonian (\ref{Hpaper}). Other asymptotic values of observables can be computed using Eqs. (\ref{constraints2}), (\ref{Trans}), and (\ref{Ademo}). Notably, the operator $\mathcal{R}$ effectively captures classical chaotic and dissipative dynamics.

In summary, we show that the classical limit of an appropriately defined MaxEnt density matrix—constructed for a general quantum-classical hybrid Hamiltonian relevant to fields such as quantum computing, quantum information, condensed matter, and quantum optics—reduces to a pure-state projector which provides the classical dynamics corresponding to the classical analogue.




\appendix
\makeatletter
\renewcommand{\@seccntformat}[1]{Appendix~\csname the#1\endcsname.\quad}
\makeatother

\section{Continuous Dependence of Solutions on Initial Conditions} \label{AppeC}
\renewcommand{\theequation}{\thesection.\arabic{equation}}
\setcounter{equation}{0}

The continuous dependence on initial conditions is analyzed through the Picard–Lindelöf Theorem~\cite{ODE}, which states that if the vector function $f(t, u)$ is continuous and satisfies a Lipschitz condition with respect to $u$, where $u$ is a vector of $n$ components, then the initial value problem

\begin{equation}
\frac{du}{dt} = f(t, u), \quad u(t_0) = u_0,
\end{equation}
has a unique solution in an interval around $u_0$. Furthermore, this theorem establishes the continuous dependence of the solutions on the initial conditions.

\subsection{Lipschitz Condition}

The Lipschitz condition is defined as

\begin{equation}
    |f(u_1) - f(u_2)| \leq L \, |u_1 - u_2|,
\end{equation}
where $L$ is a positive constant that bounds the rate of change of $f(u)$. Here, $|u|$ denotes the Euclidean norm.

We can calculate its Jacobian matrix $J_f(u)$, which contains the partial derivatives:
\begin{equation}
    J_f(u) = \left[ \frac{\partial f_i}{\partial u_j} \right].
\end{equation}
If $f(t, u)$ has continuous partial derivatives with respect to the components $u_j$, then the Lipschitz condition is fulfilled if the maximum of the absolute value of the partial derivatives is bounded by a constant that is~\cite{ODE}:
\begin{equation}
\label{desK}
\left| \frac{\partial F}{\partial u_k} \right|_M \leq L,
\end{equation}
for $k = 1, 2, \ldots, n$.

In this case, the solutions to the system of equations continuously depend on the initial conditions in $D$. In spatially bounded regions $D$, this condition always holds.

The Hamiltonians $\hat{H}_1$ and $\hat{H}_2$ satisfy this condition. Physical systems generally evolve within bounded regions; otherwise, their dynamics can be studied in defined domains.

The convergence of the solutions of the quantum-classical hybrid system (\ref{Eccanon})–(\ref{eqclasgen1}) to the classical system (\ref{class}) is ensured by taking the limit $I \rightarrow 0$ and a suitable choice of the initial conditions. For example, we can take the independent initial conditions as $\langle \hat{L} \rangle(0)$, $\langle \hat{x}^{2} \rangle(0)$, and $A(0)$ in the semiclassical case, and $L(0)$, $x^{2}(0)$, and $A(0)$ for the classical situation.

We let $\langle \hat{x}^{2} \rangle(0)$ vary in the interval

\begin{equation}
    \left( \frac{E}{\alpha_1} - \sqrt{\left( \frac{E}{\alpha_1} \right)^2 - \frac{\alpha_2}{\alpha_1} I_L}, \,
       \frac{E}{\alpha_1} + \sqrt{\left( \frac{E}{\alpha_1} \right)^2 - \frac{\alpha_2}{\alpha_1} I_L} \right),
\end{equation}
where $I_L = I + \langle L \rangle(0)^2 / 4$. This expression can be simplified by fixing $\langle L \rangle(0) = 0$ (with $I \leq (E/\alpha_1)^2$). For classical variables, we fix $L(0) = 0$ and allow $x^{2}(0)$ to vary in an equivalent interval but with $I = 0$ and variables instead of mean values.

From Eq.~(\ref{eqInv}) we see that
\begin{equation}
    \langle \hat{p}^{2} \rangle(0) = \frac{I}{\langle \hat{x}^{2} \rangle(0)},
\end{equation}
while $p^2 = 0$ in the classical instance.

In addition, from Eq.~(\ref{Hpaper}) we deduce $P_A(0)$ via
\begin{equation}
\label{PA0}
P_A = \pm  \sqrt{\frac{2E - \alpha_1 \langle \hat{x}^2 \rangle - \alpha_2 \langle \hat{p}^2 \rangle - \alpha_3 \langle \hat{L} \rangle - \beta_1 F(A) - \gamma V(A) \langle \hat{x}^{2} \rangle}{\beta_2}},
\end{equation}
evaluating all quantities at the initial time. We also have an equivalent classical version using variables instead of expectation values.

\section{Unitary Transformation} \label{AppeA}
\setcounter{equation}{0}

The operator-based functional form of the unitary transformation introduced in Ref. \cite{previo} and employed throughout this article is given by
\begin{subequations}
\label{Trans}
\begin{eqnarray}
\hat{x} & = & \frac{\sqrt{2}}{2} \left( \frac{\lambda_2}{\lambda_1} \right)^{1/4} \left[ \left( \frac{\lambda_T}{\lambda_V} \right)^{1/4} \hat{X} + \left( \frac{\lambda_V}{\lambda_T} \right)^{1/4} \hat{P} \right], \\
\hat{p} & = & \frac{\sqrt{2}}{2} \left( \frac{\lambda_1}{\lambda_2} \right)^{1/4} \left[ -\left( \frac{\lambda_T}{\lambda_V} \right)^{1/4} \hat{X} + \left( \frac{\lambda_V}{\lambda_T} \right)^{1/4} \hat{P} \right], \hspace{1cm}
\end{eqnarray}
\end{subequations}
where $\lambda_V = \sqrt{\lambda_1 \lambda_2} + \lambda_3$ and $\lambda_T = \sqrt{\lambda_1 \lambda_2} - \lambda_3$.

The operator $\hat{X}^{2} + \hat{P}^{2}$ is expressed in units of action and can be rewritten as $\hat{X}^{2}(t) + \hat{P}^{2}(t) = \hbar \left( 2 \hat{a}^{\dagger}(t) \hat{a}(t) + 1 \right)$. Consequently, the density operator $\hat{\rho}$ has eigenvalues
\begin{equation}
\label{autovalores}
\exp(-\lambda_0) \exp\left[ -\hbar I_\lambda (2n + 1) \right], \quad n = 0, 1, 2, \ldots,
\end{equation}
when expressed in the eigenbasis $|0\rangle, |1\rangle, |2\rangle, \ldots$ of $\hat{a}^{\dagger}(t) \hat{a}(t)$, which are also eigenstates of $\hat{\rho}(t)$. These eigenvalues exhibit a complex time dependence governed by the unitary transformation in Eq. (\ref{Trans}).

Since the transformation is unitary, it preserves the canonical commutation relations: $[\hat{X}, \hat{P}] = [\hat{x}, \hat{p}] = i\hbar$, and therefore $[\hat{a}^{\dagger}, \hat{a}] = 1$. From Eq. (\ref{landa0}) and Ref. \cite{previo}, it follows that
\begin{equation}
\label{landa0parti}
\lambda_0 = -\ln \left[ \exp(\hbar I_\lambda) - \exp(-\hbar I_\lambda) \right],
\end{equation}
and using Eq. (\ref{katze}), the expectation values (EVs) take the form
\begin{subequations}
\label{rel.mv,L}
\begin{eqnarray}
\langle \hat{x}^{2} \rangle & = & \frac{T(I_\lambda)}{I_\lambda} \lambda_2, \label{rel.mv4,xcuad} \\
\langle \hat{p}^{2} \rangle & = & \frac{T(I_\lambda)}{I_\lambda} \lambda_1, \\
\langle \hat{L} \rangle & = & -2 \frac{T(I_\lambda)}{I_\lambda} \lambda_3, \label{rel.mv4,L4}
\end{eqnarray}
\end{subequations}
\noindent
where $T(I_\lambda)$ is defined in Ref. \cite{previo} as
\begin{equation}
\label{tilambda}
T(I_\lambda) = \frac{\hbar}{2} \left( \frac{\exp(2\hbar I_\lambda) + 1}{\exp(2\hbar I_\lambda) - 1} \right).
\end{equation}
Moreover, from Eq. (\ref{rel.mv,L}) we deduce that
\begin{equation}
\label{I, T(Ilamba)}
T(I_\lambda) = \sqrt{I}.
\end{equation}

\bibliographystyle{elsarticle-num}

\bibliography{refences}

@PREAMBLE{
 "\providecommand{\noopsort}[1]{}" 
 # "\providecommand{\singleletter}[1]{#1}%" 
}

@article{Bloch,
  title={Nuclear induction},
  author={Bloch, Felix},
  journal={Physical review},
  volume={70},
  number={7-8},
  pages={460},
  year={1946},
  publisher={APS}
}

@book{Milonni,
  title={Chaos in laser-matter interactions},
  author={Milonni, Peter W and Shih, ML and Ackerhalt, Jay R},
  volume={6},
  year={1987},
  publisher={World Scientific Publishing Company}
}

@book{meystre07,
  title={Elements of quantum optics},
  author={Meystre, Pierre and Sargent, Murray},
  year={2007},
  publisher={Springer Science \& Business Media}
}

@book{Ring,
  title={The nuclear many-body problem},
  author={Ring, Peter and Schuck, Peter},
  year={2004},
  publisher={Springer Science \& Business Media}
}

@article{k95,
  title={Semiclassical model for quantum dissipation},
  author={Kowalski, AM and Plastino, A and Proto, AN},
  journal={Physical Review E},
  volume={52},
  number={1},
  pages={165},
  year={1995},
  publisher={APS}
}

@article{1,
  title={Bandt Pompe's Purity Description of the Quantum-Classic Transition},
  author={Plastino, Angelo and Kowalski, AM},
  journal={Available at SSRN 4575280}
}

@article{cooper94,
  title={Semiquantum chaos},
  author={Cooper, Fred and Dawson, John F and Meredith, Dawn and Shepard, Harvey},
  journal={Physical review letters},
  volume={72},
  number={9},
  pages={1337},
  year={1994},
  publisher={APS}
}

@article{bonilla92,
  title={Collapse of the wave packet and chaos in a model with classical and quantum degrees of freedom},
  author={Bonilla, LL and Guinea, F},
  journal={Physical Review A},
  volume={45},
  number={11},
  pages={7718},
  year={1992},
  publisher={APS}
}

@article{patta92,
  title={Effective potentials and chaos in quantum systems},
  author={Pattanayak, Arjendu K and Schieve, William C},
  journal={Physical Review A},
  volume={46},
  number={4},
  pages={1821},
  year={1992},
  publisher={APS}
}

@article{Fratino14,
  title={Entanglement dynamics in a quantum--classical hybrid of two q-bits and one oscillator},
  author={Fratino, Lorenzo and Lampo, Aniello and Elze, Hans Thomas},
  journal={Physica Scripta},
  volume={2014},
  number={T163},
  pages={014005},
  year={2014},
  publisher={IOP Publishing}
}

@article{Alonso20,
  title={Entropy and canonical ensemble of hybrid quantum classical systems},
  author={Alonso, JL and Bouthelier, C and Castro, A and Clemente-Gallardo, J and Jover-Galtier, JA},
  journal={Physical Review E},
  volume={102},
  number={4},
  pages={042118},
  year={2020},
  publisher={APS}
}

@article{boghiu24,
  title={Hybrid quantum-classical control problems},
  author={Boghiu, Emanuel-Cristian and Mart{\'\i}nez-Crespo, David and Jover-Galtier, Jorge A and Clemente-Gallardo, Jes{\'u}s},
journal={Communications in Analysis and Mechanics},
volume={16},
  year={2024},
publisher={AIMS Press}
}

@article{3,
  title={Decoherence, einselection, and the quantum origins of the classical},
  author={Zurek, Wojciech Hubert},
  journal={Reviews of modern physics},
  volume={75},
  number={3},
  pages={715},
  year={2003},
  publisher={APS}
}

@article{k19renyi,
  title={A nonlinear matter-field Hamiltonian analyzed with Renyi and Tsallis statistics},
  author={Kowalski, AM and Plastino, A},
  journal={Physica A: Statistical Mechanics and its Applications},
  volume={535},
  pages={122387},
  year={2019},
  publisher={Elsevier}
}

@article{k11fisher,
  title={Fisher information description of the classical--quantal transition},
  author={Kowalski, Andres Mauricio and Mart{\'\i}n, Mar{\'\i}a Teresa and Plastino, A and Rosso, Osvaldo Anibal},
  journal={Physica A: Statistical Mechanics and its Applications},
  volume={390},
  number={12},
  pages={2435--2441},
  year={2011},
  publisher={Elsevier}
}

@book{4,
  title={Semiclassical mechanics with molecular applications},
  author={Child, Mark Sheard},
  year={2014},
  publisher={Oxford University Press, USA}
}

@article{Kowalski2002,
  title={Classical limits},
  author={Kowalski, Andres M and Plastino, A and Proto, AN},
  journal={Physics Letters A},
  volume={297},
  number={3-4},
  pages={162--172},
  year={2002},
  publisher={Elsevier}
}

@article{BP,
  title={Bandt--Pompe approach to the classical-quantum transition},
  author={Kowalski, AM and Mart{\'\i}n, MT and Plastino, A and Rosso, OA},
  journal={Physica D: Nonlinear Phenomena},
  volume={233},
  number={1},
  pages={21--31},
  year={2007},
  publisher={Elsevier}
}

@article{energia,
  title={Classical Limit, Quantum Border and Energy},
  author={Kowalski, Andres Mauricio and Plastino, Angelo and Gonzalez, Gaspar},
  journal={Physics},
  volume={5},
  number={3},
  pages={832--850},
  year={2023},
  publisher={MDPI}
}

@article{previo,
  title={Chaotic density matrix in the classical limit},
  author={Kowalski, AM and Plastino, A},
  journal={Physics Letters A},
  volume={384},
  number={24},
  pages={126450},
  year={2020},
  publisher={Elsevier}
}

@article{caosclasico,
  title={Classical chaos described by a Density matrix},
  author={Kowalski, Andres Mauricio and Plastino, Angelo and Gonzalez, Gaspar},
  journal={Physics},
  volume={3},
  number={3},
  pages={739--746},
  year={2021},
  publisher={MDPI}
}

@article{disipativo,
  title={Dynamical classic limit: Dissipative vs conservative systems},
  author={Gonzalez Acosta, G and Plastino, A and Kowalski, AM},
  journal={Chaos: An Interdisciplinary Journal of Nonlinear Science},
  volume={33},
  number={1},
  year={2023},
  publisher={AIP Publishing}
}

@article{Katz,
  title={Principles of statistical mechanics: the information theory approach},
  author={Katz, Amnon},
  journal={W. H. Freeman and Company},
  year={1967}
}

@article{Levine,
  title={Connection between the maximal entropy and the scattering theoretic analyses of collision processes},
  author={Alhassid, Y and Levine, RD},
  journal={Physical Review A},
  volume={18},
  number={1},
  pages={89},
  year={1978},
  publisher={APS}
}

@misc{ODE,
  title={Theory of ordinary differential equations},
  author={Coddington, Earl A and Levinson, Norman and Teichmann, T},
  year={1956},
  publisher={American Institute of Physics}
}

@article{gonzalez2023mixed,
  title={Mixed quantum-semiclassical simulation},
  author={Gonzalez-Conde, Javier and Sornborger, Andrew T},
  journal={arXiv preprint arXiv:2308.16147},
  year={2023}
}

@article{kurizki2015quantum,
  title={Quantum technologies with hybrid systems},
  author={Kurizki, Gershon and Bertet, Patrice and Kubo, Yuimaru and M{\o}lmer, Klaus and Petrosyan, David and Rabl, Peter and Schmiedmayer, J{\"o}rg},
  journal={Proceedings of the National Academy of Sciences},
  volume={112},
  number={13},
  pages={3866--3873},
  year={2015},
  publisher={National Academy of Sciences}
}

@article{yu06nanotec,
  title={Nanotechnology: Role in emerging nanoelectronics},
  author={Yu, B and Meyyappan, M},
  journal={Solid-state electronics},
  volume={50},
  number={4},
  pages={536--544},
  year={2006},
  publisher={Elsevier}
}

@article{das10meso,
  title={Mesoscopic systems in the quantum realm: fundamental science and applications},
  author={Das, Mukunda P},
  journal={Advances in Natural Sciences: Nanoscience and Nanotechnology},
  volume={1},
  number={4},
  pages={043001},
  year={2010},
  publisher={IOP Publishing}
}

\end{document}